\documentclass[pre]{revtex4}

\usepackage{amsmath}
\usepackage{amsfonts}
\usepackage{amssymb}
\usepackage{graphicx}

\renewcommand{\natural}{{\vrule height .48em width .03em depth 0ex
\kern -.30em {\rm N}}}

\DeclareGraphicsExtensions{.eps, .ps}

\begin{document}

\title{Hawks and Doves on Small-World Networks}

\author{Marco Tomassini}
\email{marco.tomassini@unil.ch}
\affiliation{Information Systems Institute, HEC, University of Lausanne, Switzerland}

\author{Leslie Luthi}
\email{leslie.luthi@unil.ch}
\affiliation{Information Systems Institute, HEC, University of Lausanne, Switzerland}

\author{Mario Giacobini}
\email{mario.giacobini@unil.ch}
\affiliation{Information Systems Institute, HEC, University of Lausanne, Switzerland}

\begin{abstract}
We explore the Hawk-Dove game on networks with topologies ranging from regular lattices to
random graphs with small-world networks in between.
This is done by means of computer simulations using several update rules for the population evolutionary dynamics.
We find the overall result that cooperation is sometimes inhibited and sometimes enhanced in
those network structures, with respect to the mixing population case.
The differences are due to different update rules and depend on the gain-to-cost ratio. 
We analyse and qualitatively explain this behavior by using local topological arguments.
\end{abstract}

\maketitle

\section{\label{intro}Introduction}
Hawks and Doves, also known as Chicken, or the Snowdrift game, is a two-person, symmetric
game with the following payoff bi-matrix:

\begin{table}[!ht]
\begin{center}
{\normalsize
\begin{tabular}{c|cc}
 & H & D\\
\hline
{\rule[-3mm]{0mm}{8mm}}
H & ($\frac{G-C}{2},\frac{G-C}{2}$) & ($G,0$)\\
{\rule[-3mm]{0mm}{4mm}}
D & ($0,G$) & ($\frac{G}{2},\frac{G}{2}$)\\
\end{tabular}
}
\end{center}
\end{table}
\vspace{-0.4cm}

\noindent In this matrix, H stands for ``hawk'', and D stands for ``dove''.
Metaphorically, a hawkish behavior means a strategy of fighting, while a dove, when facing a confrontation, will always yield.
As in the 
\textit{Prisoner's Dilemma} \cite{axe84}, this game, for all its simplicity, appears to
capture some important features of  social interactions. In this sense, it applies
in many situations in which ``parading'', ``retreating'', and ``escalading'' are common.
One striking example of a situation
that has been thought to lead to a Hawk-Dove dilemma is the Cuban missile crisis in 1962
 \cite{poundstone92}.
In the payoff matrix above, $G > 0$ is the gain that a hawk obtains when it meets a dove; the dove retreats and looses nothing.
If a dove meets another dove, one of them, or both, will retreat and they will gain half of
the price each ($G/2$) in the average. Finally, when a hawk meets another hawk, they 
both fight and each obtains an average payoff of $(G-C)/2$, where $C$ is the cost of any
injury that might occur in the fight. It is assumed that $C > G$, i.e. the cost of injury
always exceeds the prize of the fight.
The game has the same structure as the Prisoner's Dilemma in that if both players
cooperate (i.e. they play dove), they both gain something, although there is a strong motivation
to act aggressively (i.e. to play the hawk strategy). However, in this game one makes the
assumption that one player is willing to cooperate, even if the other does not, and that mutual defection, i.e. result (H,H), is detrimental to both players. 

In contrast to the Prisoner's Dilemma which has a unique Nash equilibrium that corresponds to
both players defecting, the Hawk-Dove game has two Nash equilibria in pure strategies
(H,D) and (D,H), and a third equilibrium in mixed strategies where strategy H is played
with probability $G/C$, and strategy D with probability $1 - G/C$. Note that we only consider
one-shot games in this work; repeated games are not taken into account.

Considering now not just two players but rather a large, mixing population of identical players,
\textit{evolutionary
game theory} \cite{hofb-sigm-book-98} prescribes that the only evolutionarily
stable strategy (ESS) of the population is the mixed strategy, giving rise, at equilibrium,
to a frequency of hawks in the population equal to $G/C$.
In the case of the Prisoner's Dilemma, one finds a unique ESS with all the individuals defecting.
However,
in 1992, Nowak and May \cite{nowakmay92} showed
that cooperation in the population is sustainable in the Prisoner's Dilemma under certain conditions,
provided that the network of the interactions between players has a lattice spatial structure. Killingback and Doebeli \cite{KD-96} extended the
spatial approach to the Hawk-Dove game and found that a planar lattice structure
with only nearest-neighbor interactions may favor cooperation, i.e. the fraction of doves in
the population is often higher than what is predicted by evolutionary game theory. In addition,
complex dynamics resembling phase transitions were observed, which is not the case in the
mixing population.
In a more recent work however, Hauert and Doebeli  \cite{hauer-doeb-2004} were led to a different conclusion, namely that
spatial structure does not seem to favor cooperation in the Hawk-Dove game. 
Additional
results on the Hawk-Dove game on a two-dimensional lattice have been recently obtained by Sysi-Aho et al.
\cite{myopic-hd-05} using a simple local decision rule for the players that does not reduce to the customary
\textit{replicator} or \textit{imitation} dynamics \cite{hofb-sigm-book-98}. They concluded that,
with their player's decision rule, cooperation persists, giving results different from those obtained
with the replicator dynamics. 
These apparently
contradictory results aroused our curiosity, and motivated us to study the situation in a more general
setting, in which the mixing population and the lattice are special cases. 

Following pioneering work by sociologists in the sixties such as that of Milgram
\cite{milgram67}, in the last few years it has become apparent that the topological structures of social
interactions networks have particular, and partly unexpected, properties that are a consequence
of their \textit{small-world} characteristics. Roughly speaking, small-world networks are
graphs that have a short \textit{average path length}, i.e. any node is relatively close to any other
node, like random graphs and unlike regular lattices.
However, in contrast with random graphs, they also have a certain amount of local structure,
as measured, for instance, by a quantity called the \textit{clustering coefficient} (an excellent
review of the subject is \cite{newman-03}).
In the same vein, many real conflicting situations in economy and sociology are not well 
described neither by a fixed
geographical position of the players in a regular lattice, nor by a mixing population or a
random graph. Starting from the two limiting cases of a random-graph and the two-dimensional lattice, our objective
here is to study the Hawk-Dove game on small-world networks in order to cover the ``middle ground''
between them. Although the Watts--Strogatz networks \cite{watts-strogatz-98} used here are not faithful representations
of the structure of real social networks, they are a useful first step toward a better understanding
of evolutionary games on networks. While we study here the Hawk-Dove game, this class of networks has been previously used for the Prisoner's 
Dilemma in \cite{social-pd-kup-01,pd-dyn-sw-02,watts99}. The work of \cite{social-pd-kup-01} is especially relevant
for our present study, while the two others deal either with special features such as
``influential individuals''\cite{pd-dyn-sw-02}, or refer to iterated versions of the game \cite{watts99}.

Recently, Santos and Pacheco \cite{santos-pach-05} have investigated both the Prisoner's
Dilemma and Hawk-Dove games on
fixed scale-free networks. The main observation from their simulations is that, at least on preferential attachment
networks, the amount of cooperative behavior is much higher than in either mixing or 
lattice-structured populations. In the abstract, and in some particular social situation
in which some individuals have an unusually high number of contacts than the rest, this is an interesting result.
However, scale-free graphs, which characterize the web and Internet among others, are not a realistic 
model of most observed social networks for various reasons (see \cite{jin-gir-newman-01,ebel-dav-born-03}),
which is why we do not comment further on the issue.

\section{\label{sect:model}The Model}
In this section we present our network models and their dynamical properties.

\subsection{\label{pop-topo}Population Topologies}
We consider a population $P$ of $N$ players where
each individual $i$ in $P$ is represented as a vertex $v_i$  of a graph $G(V,E)$,
with $v_i \in V, \; \forall i \in P$. An interaction between two players $i$ and
$j$ is represented by the undirected edge $e_{ij} \in E, \; e_{ij} \equiv e_{ij}$.
The number of neighbors of player $i$ is the degree $k_i$ of vertex $v_i$. The average
degree of the network will be called $\bar k$.

We shall use three main graph population structures: regular lattices, random graphs, and small-worlds.
In fact, our goal is to explore significant population network structures that somehow fall
between the regular lattice and random graph limits, including the bounding cases.

Our regular lattices are two-dimensional with $k_i=8, \; \forall v_i \in V$  and periodic boundary conditions. 
This neighborhood is usually called the Moore neighborhood and comprises nine individuals, including the central node.
We would like to stress that we believe regular lattice structures are not realistic representations
of most actual population structures, especially human, except when mobility and dispersal ability
of the individuals are limited as, for example, in plant ecology and territorial animals. 
The main reasons why lattices have been so heavily used is that they are more amenable to mathematical analysis and are easier to simulate.
We include them here for two reasons: as an interesting bounding case, and to allow comparison with previous work.
 
The small-world networks used here are similar to the graphs proposed by Watts and Strogatz
\cite{watts-strogatz-98}. However, there are two main differences (see \cite{boccara-04}). First, we start from a 
two-dimensional regular lattice substrate, instead of a one-dimensional lattice. This does not
modify the main features of the resulting graphs, as observed in \cite{watts99}, and as measured by us.
The reason for starting from a two-dimensional lattice is to keep with the customary ordered population
topology that is used in structured evolutionary games.  Although
they have been used as a starting point for the Prisoner's Dilemma by Abramson and Kuperman \cite{social-pd-kup-01}, one-dimensional lattices do not
make much sense in a social or biological setting, although after some rewiring the effect of
the substrate becomes almost negligible. 

The second difference is in the rewiring process. The
algorithm used here comes from \cite{boccara-04} and works as follows: starting from a regular two-dimensional lattice with
periodic boundary conditions, visit each edge and, with probability $p$, replace it  by
an edge between two randomly selected vertices, with the constraint that two vertices are
not allowed to be connected by more than one edge. As in the original Watts--Strogatz model,
the average vertex degree $\bar k$ does not change, and the process may produce disconnected graphs, 
which have been avoided in our simulations.
The advantage of this construction is that, for $p \rightarrow 1$,
the graph approaches a classical Erd\"os--R\'enyi random graph, while this is not the case for
the original Watts--Strogatz construction, since in the latter, the degree of any vertex is 
always larger than or
equal to $k/2$, $k$ being the degree of a vertex in the original lattice. 

We would like to point out that it is known that Watts--Strogatz small worlds 
are not adequate representations of social networks. Although they share some common statistical
properties with the latter, i.e. high clustering and short average path length, they lack other features that characterize
real social networks such as clusters, and dynamical self-organization \cite{ebel-dav-born-03}. In 
spite of these shortcomings, they are
a convenient first approximation for studying the behavior of agents in situations where the interaction
network is neither regular, nor random. Note also that once fixed, the interaction network
does not change during the system evolution in our study, only the strategies may evolve. Evolutionary games
on dynamic networks have been studied, for instance in \cite{zimm-et-al-04,luthi-giac-tom-05,games-ecal-05}.

\subsection{\label{dyn}Population Dynamics}
\subsubsection{Local Dynamics}
The local dynamics of a player $i$ only depends on its own strategy $s_i \in \{H,D\}$, and on
the strategies of the $k_i$ players in its neighborhood $N(i)$. Let us call $M$ the payoff matrix
of the game (see section \ref{intro}). The quantity
$$G_i(t) = \frac{1}{k_i} \sum _{j \in N(i)} s_i(t)\; M\; s_{j}^T(t)$$
\noindent is the average payoff collected by player $i$ at time step $t$.
Note that in our study, $i \notin N(i)$ meaning that self-interaction is not considered when
calculating the average payoff of an individual.
Self-interaction has  traditionally been taken into account in some previous work on the
Prisoner's Dilemma
game on grids \cite{nowakmay92,nowaketal94} on the grounds that, in biological applications,
several entities may occupy a single patch in the network. Nowak et al. find that
self-interaction does not qualitatively change the results in the Prisoner's Dilemma game. In the Hawk-Dove game
self-interaction is usually not considered; moreover, in this work we wish to compare results
with those of \cite{KD-96,hauer-doeb-2004}, where self-interaction is not included.

We use three types of rules to update a player's strategy. The rules are among those employed by
Hauert et al. \cite{hauer-doeb-2004} to allow for comparison of the results in regular lattices
and in small-world networks. Decision rules based on the player's satisfaction degree, 
such as those used in \cite{zimm-et-al-04,games-ecal-05,luthi-giac-tom-05,myopic-hd-05} are not examined here.
The rules are the following:
\begin{enumerate}
\item replicator dynamics;
\item proportional updating;
\item best-takes-over.
\end{enumerate}

The \textit{replicator dynamics} rule aims at maximal consistency with the original
evolutionary game theory equations. Player $i$ is updated by drawing
another player $j$ at random from the neighborhood $N(i)$
and replacing $s_i$ by $s_j$ with probability $p_j = \phi(G_j - G_i)$ \cite{hofb-sigm-book-98}.

The \textit{proportional updating} rule is also a stochastic rule. All the players in the neighborhood $N(i)$,
plus the player $i$ itself compete for the strategy $i$ will take at the next time step, each with a probability $p_j$ given by
$$ p_j = \frac{G_j}{\sum_l G_l}, \;\; l,j \in \{N(i) \cup i \}.$$
\noindent Negative payoffs cannot be used with this rule, since the probabilities of replication
must be $p_j \ge 0$. In order to avoid negative, or zero, values, the payoffs have been
shifted by an amount equal to the cost $C$ which, of course, leaves the game's Nash equilibria invariant.

In \textit{best-takes-over}, the strategy $s_i(t)$ of individual $i$ at time step $t$ will
be
$$s_i(t) = s_j(t-1),$$
where
$$j \in \{N(i) \cup i\} \;s.t.\; G_j = \max \{G_k(t-1)\}, \; \forall k \in \{N(i) \cup i\}.$$
\noindent That is, individual $i$ will adopt the strategy of the player with the highest
payoff among its neighbors.
If there is a tie, the winner individual is chosen uniformly at random between the best, and its strategy
replaces the current strategy of player $i$, otherwise the rule is deterministic. It should be
noted that this rule does not fit to the usual continuous evolutionary game theory which
leads to replicator dynamics, since the update
decision is a step function.

\subsubsection{Global Dynamics}
Calling $C(t) = (s_1(t), s_2(t), \ldots , s_N(t))$ a \textit{configuration} of the population
strategies at time
step $t$, the global \textit{synchronous} system dynamics is implicitly given by:
$$C(t) = F(C(t-1)), \;\; t =1,2, \ldots $$
\noindent where $F$ is the evolution operator.

Synchronous update, with its idealization of a global clock, is customary 
in spatial evolutionary games, and most results have been obtained using this model 
\cite{nowakmay92,KD-96}.
However, perfect synchronicity is only an abstraction. Indeed, in some biological
and, particularly, sociological environments, agents normally act at different and possibly uncorrelated
times, which seems to preclude a faithful globally synchronous simulation in most
cases of interest \cite{hubglance93}. In spite of this, it has been shown that the
update mode does not fundamentally alter the results, as far
as evolutionary games are concerned \cite{nowaketal94,hauer-doeb-2004}. In this paper we
present results for both synchronous and asynchronous dynamics.

Asynchronous dynamics must nevertheless be further qualified, since there 
are many ways for serially updating the strategies
of the agents. Here we use the discrete update dynamics that makes the least assumption
about the update sequence: the next cell to be updated is chosen
at random with uniform probability and with replacement.
This corresponds to a binomial distribution of the updating probability and is a good approximation of a continuous-time Poisson
process. This asynchronous update is analogous to the one used by Hauert et al. 
\cite{hauer-doeb-2004}, which will allow us to make meaningful comparisons.

\section{\label{sim}Simulation Results}
In order to analyze the influence of the structure of the network
on the proportion of cooperation (i.e. dove behavior),
2500 players were organized into 5 different networks:
a 50 by 50 toroidal lattice where every cell is connected to its 8 nearest neighbors,
three different small-world networks, and the random graph.
The three categories of small worlds are obtained by rewiring each edge
with a certain probability $p$ using the technique described under \ref{pop-topo}.
The values used are $p \in \{0.01,0.05,0.1\}$.
The random graph is generated by first creating the lattice
and then rewiring each link, in the same manner used to construct
small worlds, but with probability $p=1$. Although our population size is smaller
than that used in \cite{hauer-doeb-2004}, which is $10000$, results turn out to be
 qualitatively similar and comparable.
For each of the 5 networks mentioned above and for all update policies,
50 runs of 5000 time steps each were executed.
In the following figures, the curves indicating the proportion of doves in the population
were obtained by averaging over the last 10 time steps of each run, well after all transients
have decayed.
At the beginning of each run, we generate a new network of the type being studied
and randomly initialize it with 50\% doves and 50\% hawks. For completeness, we mention
that experiments with 10\% and 90\% initial cooperators respectively, give results that
are qualitatively indistinguishable from the 50\% case in the long run. Therefore, we do not include the 
corresponding graphs for reasons of space.

In the following figures, the dashed diagonal line going from a fraction of cooperators
of $1$ for $r=0$, to a fraction of $0$ for $r=1$, represents the equation $1-G/C = 1-r$,
which is the equilibrium fraction of cooperators as a function of $r$ given by the standard replicator-dynamics equations \cite{hofb-sigm-book-98},
and it is reported here for the sake of comparison. 
It should be noted, however, that the simulations are not expected to fit this line. The reason is
that the analytic solution is obtained under two main hypotheses: the population size is very
large, and individuals are matched pairwise randomly. These conditions are not satisfied by
the finite-size, discrete systems used for the simulations, and thus one should not expect strict
adherence to the mean-field equations. On the other hand, the type of \textit{mesoscopic} system
simulated here is probably closer to reality, where finiteness and discreteness are the rule.
Another reason why we do not expect the results of the simulations to closely fit the
theoretical solution is that two of the local update rules
(best-takes-over and proportional updating) do not reduce to the standard replicator dynamics.

This section is subdivided into three separate parts, one per decision rule previously
mentioned under \ref{dyn}.

\subsection{Replicator Dynamics}
To determine the probability $p_j$ for replacing an individual $i$, having a gain $G_i$, 
by one of its randomly chosen neighbors $j$, whose gain is $G_j$,
we use the the previously introduced function $\phi(G_j - G_i)$ as follows:
\begin{eqnarray}
p_j = \phi(G_j - G_x) =
\begin{cases} \textrm{\large{$\frac{G_j - G_i}{{d}_{max}}$}} & \textrm{if $G_j - G_i > 0$}\\\\
0 & \textrm{otherwise}
\end{cases}
\label{repl_dyn_eq}
\end{eqnarray}
where ${d}_{max} = \frac{G+C}{2}$ is the largest difference in gain
there can be between two players.

With this definition of $\phi$, individual $i$ imitates neighbor $j$'s strategy
with a certain probability proportional to the difference of their average payoffs
and only if $j$ has a higher gain than $i$.
Notice that if $i$ and $j$ have the same average payoffs, $i$'s strategy is left untouched, while
if $G_j -G_i = {d}_{max}$, $i$ necessarily adopts $j$'s strategy.

\begin{figure}[!ht]
\begin{center}
\begin{tabular}{ccc}
\mbox{\includegraphics[width=5.5cm, height=5.5cm]{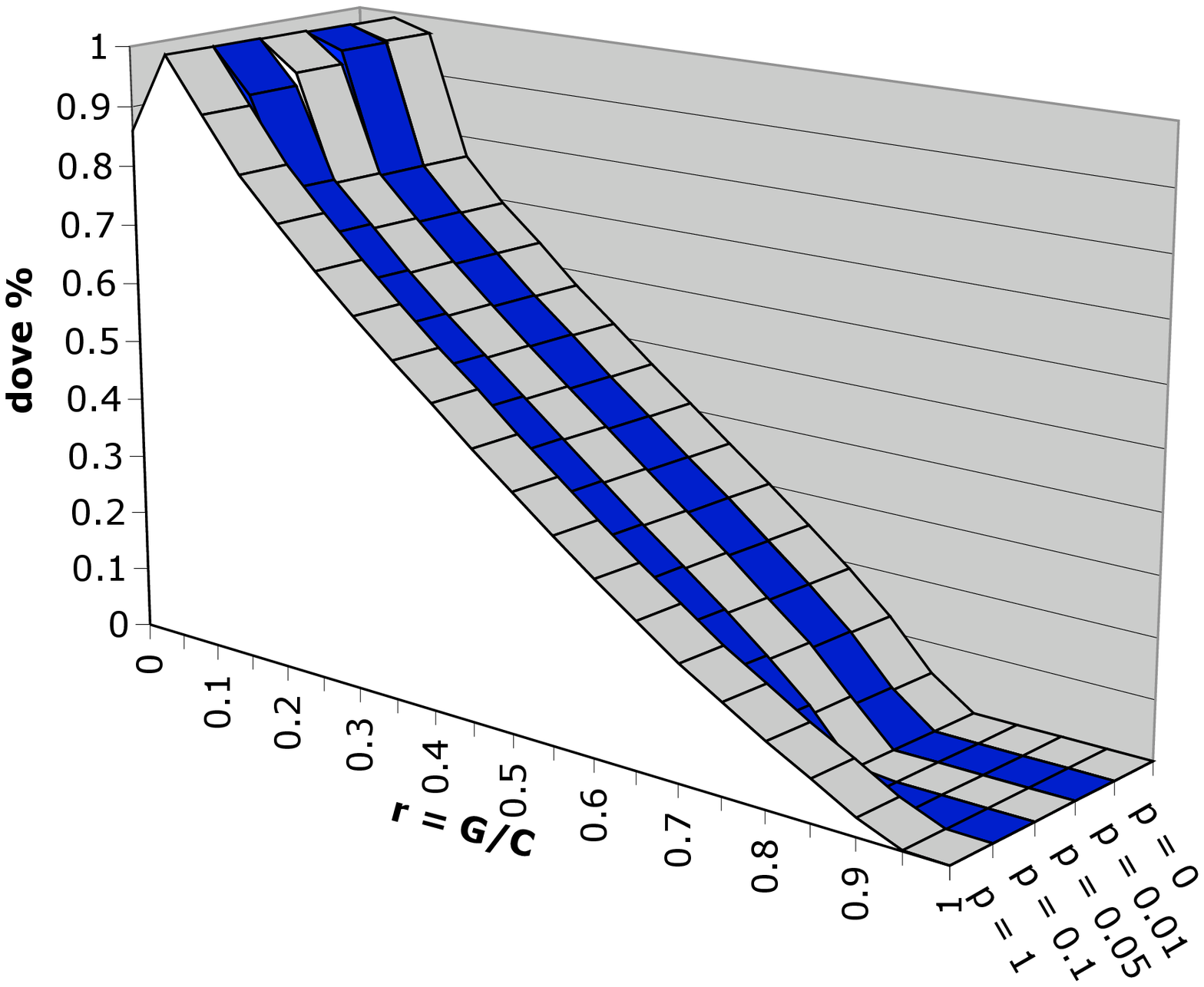}}\protect & \hspace*{1cm} &
\mbox{\includegraphics[width=5.5cm, height=5.5cm]{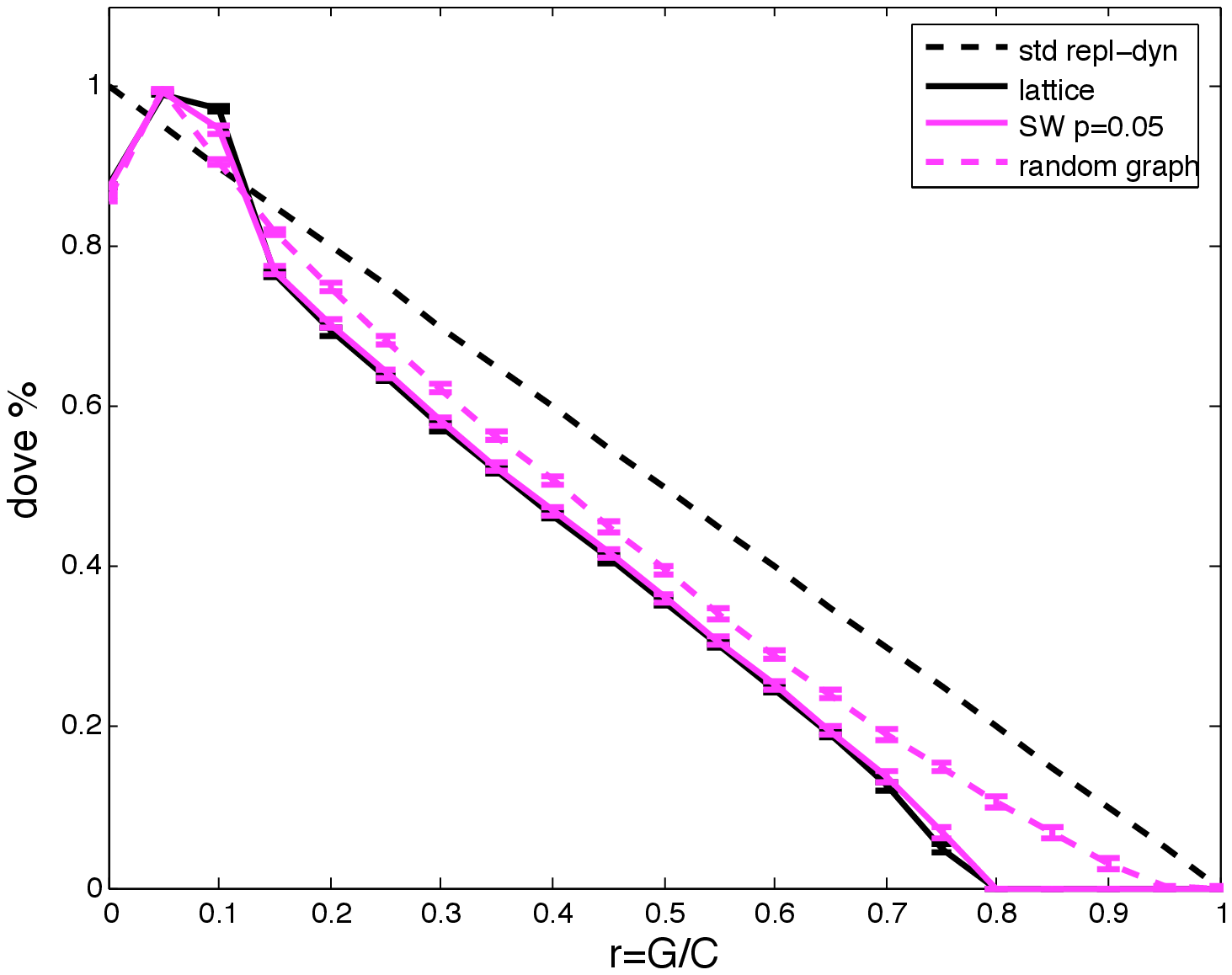}}\\
(a) & &(b)
\end{tabular}
\end{center}
\caption{\label{repl_dyn_async}(Color online) asynchronous replicator dynamics updating;
(a) frequency of doves as a function of the gain-to-cost ratio $r$ for differents topologies:
lattice ($p=0$), small worlds ($p=0.01$, $p=0.05$, $p=0.1$), random graph ($p=1$);
(b) small world with $p = 0.05$ compared to the grid ($p=0$) and random graph ($p=1$) cases.
Bars indicate standard deviations and the diagonal dashed line is $1-r$ (see text).}
\end{figure}

\begin{figure}[!ht]
\begin{center}
\begin{tabular}{ccc}
\mbox{\includegraphics[width=5.5cm, height=5.5cm]{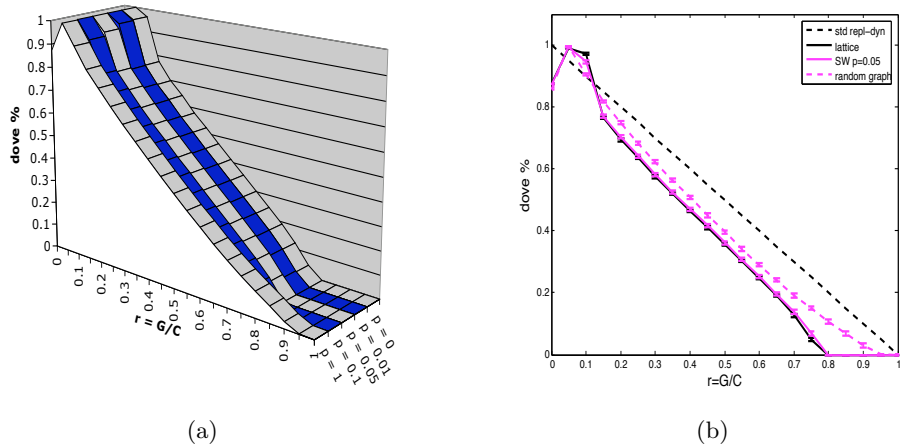}}\protect & \hspace*{1cm} &
\mbox{\includegraphics[width=5.5cm, height=5.5cm]{repl_dyn_async_uc_dPercentage}}\\
(a) & & (b)
\end{tabular}
\end{center}
\caption{\label{repl_dyn_sync}(Color online) synchronous replicator dynamics updating;
(a) frequency of doves as a function of the gain-to-cost ratio $r$ for differents topologies:
lattice ($p=0$), small worlds ($p=0.01$, $p=0.05$, $p=0.1$), random graph ($p=1$);
(b) small world with $p = 0.05$ compared to the grid ($p=0$) and random graph ($p=1$) cases.
Bars indicate standard deviations and the diagonal dashed line is $1-r$ (see text).}
\end{figure}

Now taking a look at figures \ref{repl_dyn_async} and \ref{repl_dyn_sync},
we clearly observe that for both synchronous and asynchronous dynamics,
cooperation is globally inhibited by spatial structure, confirming the results of \cite{hauer-doeb-2004}.
Even the case of the random graph generates higher rates of hawks.
Further details as to why this may occur can be found in section \ref{disc}.

We note in passing that the experimental curve corresponding to the random graph limit appears
to be close to the curve corresponding to the pair approximation calculation in Hauert and Doebeli's
work \cite{hauer-doeb-2004}. This is not surprising, given that pair approximation works
better in random graphs than in regular lattices, unless higher-order effects are taken into
account \cite{van-baalen-00}. Since the curves for the random graphs in 
figures \ref{repl_dyn_async} and \ref{repl_dyn_sync} are averages over many graph realizations,
each pair has some probability to contribute in the simulation,
which explains the resemblance between our experimental curves and the calculations of \cite{hauer-doeb-2004}.

\subsection{Proportional Updating}
\begin{figure}[!ht]
\begin{center}
\begin{tabular}{ccc}
\mbox{\includegraphics[width=5.5cm, height=5.5cm]{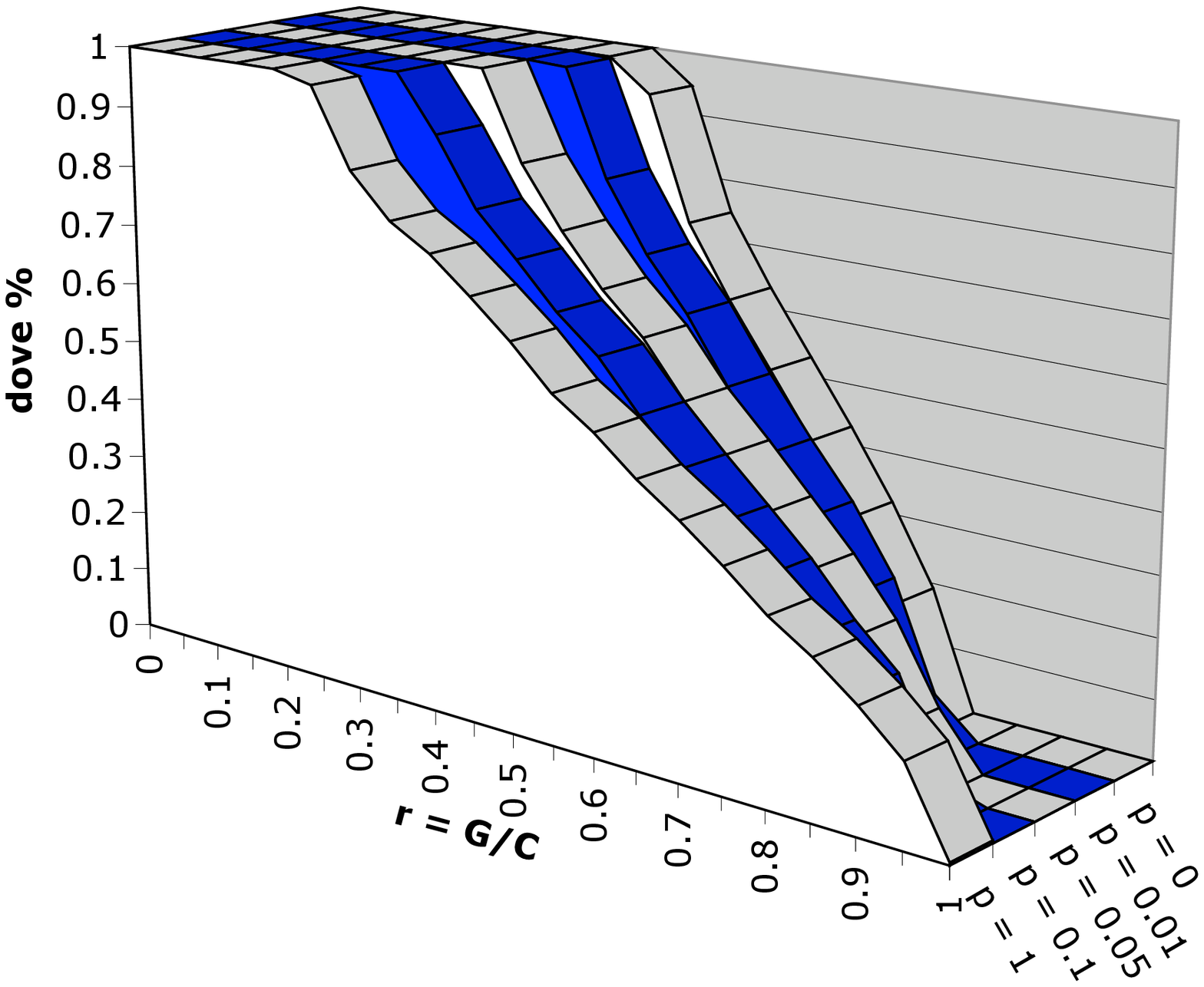}}\protect & \hspace*{1cm} &
\mbox{\includegraphics[width=5.5cm, height=5.5cm]{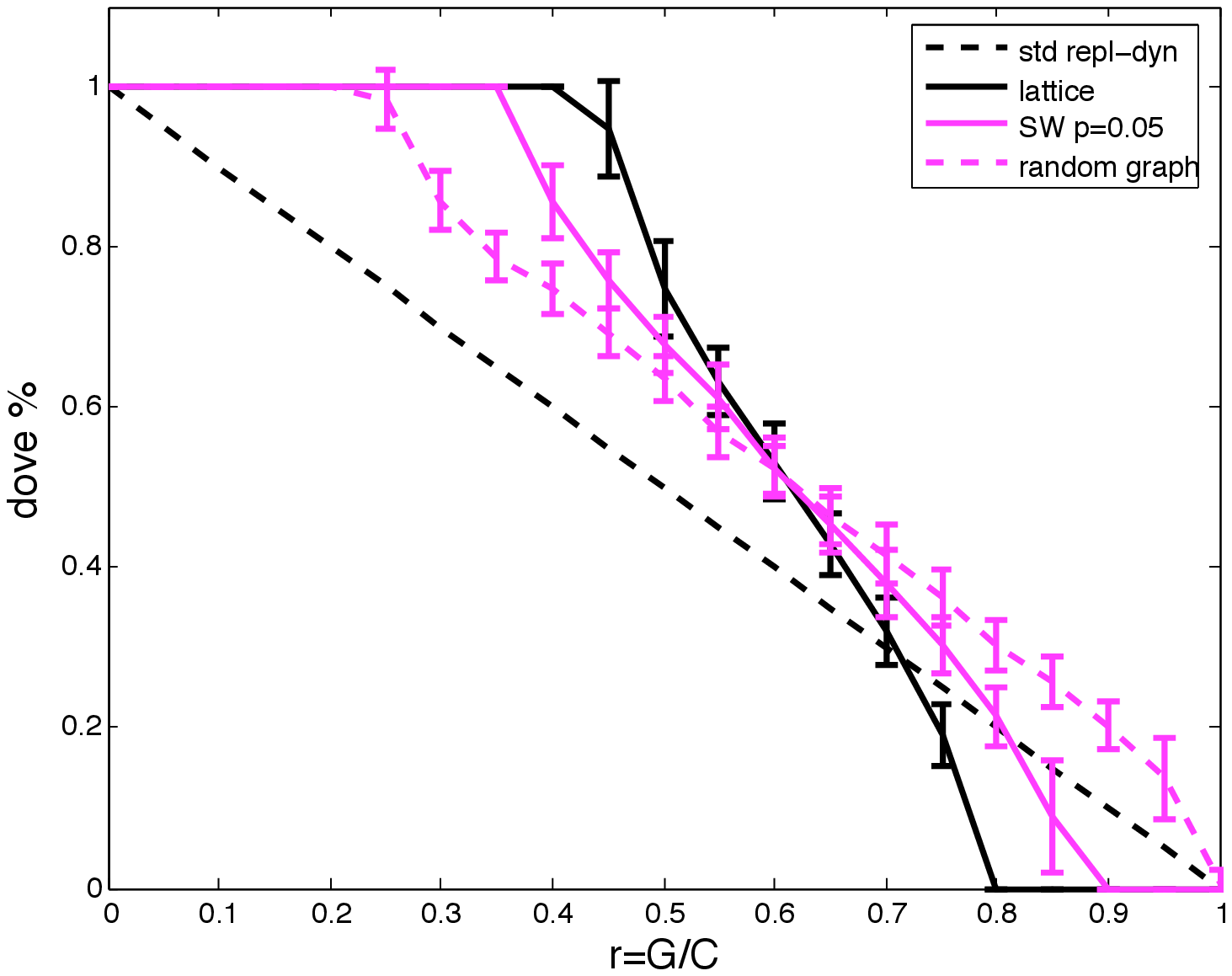}}\\
(a) & & (b)
\end{tabular}
\end{center}
\caption{\label{prop_async}(Color online) asynchronous proportional udpating;
(a) frequency of doves as a function of the gain-to-cost ratio $r$ for differents topologies:
lattice ($p=0$), small worlds ($p=0.01$, $p=0.05$, $p=0.1$), random graph ($p=1$);
(b) small world with $p = 0.05$ compared to the grid ($p=0$) and random graph ($p=1$) cases.
Bars indicate standard deviations and the diagonal dashed line is $1-r$ (see text).}
\end{figure}

\begin{figure}[!ht]
\begin{center}
\begin{tabular}{ccc}
\mbox{\includegraphics[width=5.5cm, height=5.5cm]{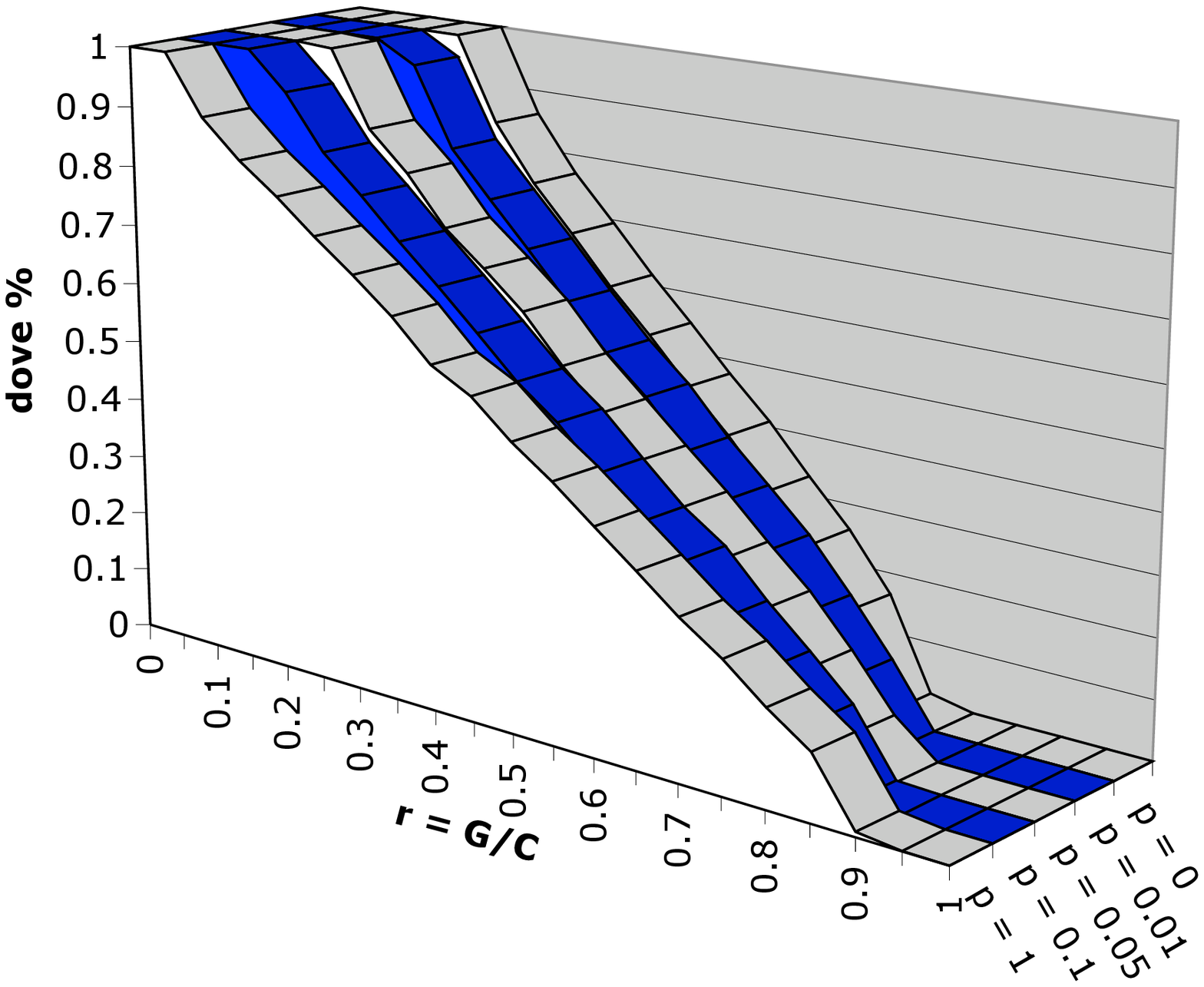}}\protect & \hspace*{1cm} &
\mbox{\includegraphics[width=5.5cm, height=5.5cm]{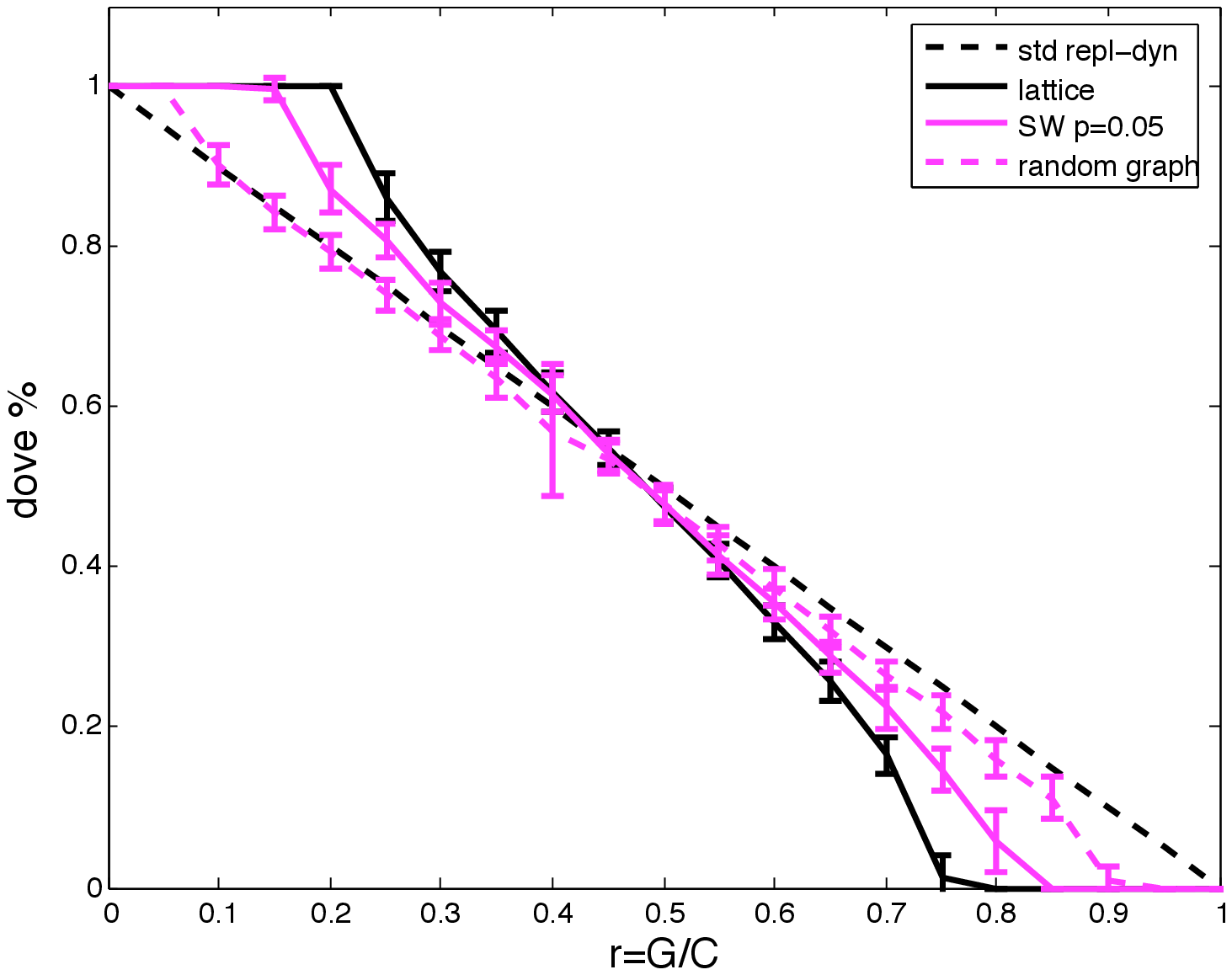}}\\
(a) & & (b)
\end{tabular}
\end{center}
\caption{\label{prop_sync}(Color online) synchronous proportional updating;
(a) frequency of doves as a function of the gain-to-cost ratio $r$ for differents topologies:
lattice ($p=0$), small worlds ($p=0.01$, $p=0.05$, $p=0.1$), random graph ($p=1$);
(b) small world with $p = 0.05$ compared to the grid ($p=0$) and random graph ($p=1$) cases.
Bars indicate standard deviations and the diagonal dashed line is $1-r$ (see text).}
\end{figure}

Figures \ref{prop_async} and \ref{prop_sync} show that,
when using the proportional updating rule,
spatial structure neither favors nor inhibits dove-like behavior
contrary to what \cite{KD-96} and \cite{hauer-doeb-2004} seem to suggest.
Indeed, for low values of $r$, the more the network is structured,
the higher the proportion of doves.
However as $r$ increases, the tendency is reversed,
thus giving a lower percentage of doves in the lattice and small-world networks
than present in the random graph topology.
This phenomenon is even more marked when using the asynchronous update.

Thus when using the proportional updating rule, if spatial structure should favor one strategy over the other for a given value of $r$,
it would be the one that is already present in greater numbers
when the topology is a random graph.

Another interesting aspect observed is the higher percentage of doves
when updating asynchronously compared to the synchronous equivalent.
This will be discussed in more detail in section \ref{disc}.

\subsection{Best-takes-over}
\begin{figure}[!ht]
\begin{center}
\begin{tabular}{ccc}
\mbox{\includegraphics[width=5.5cm, height=5.5cm]{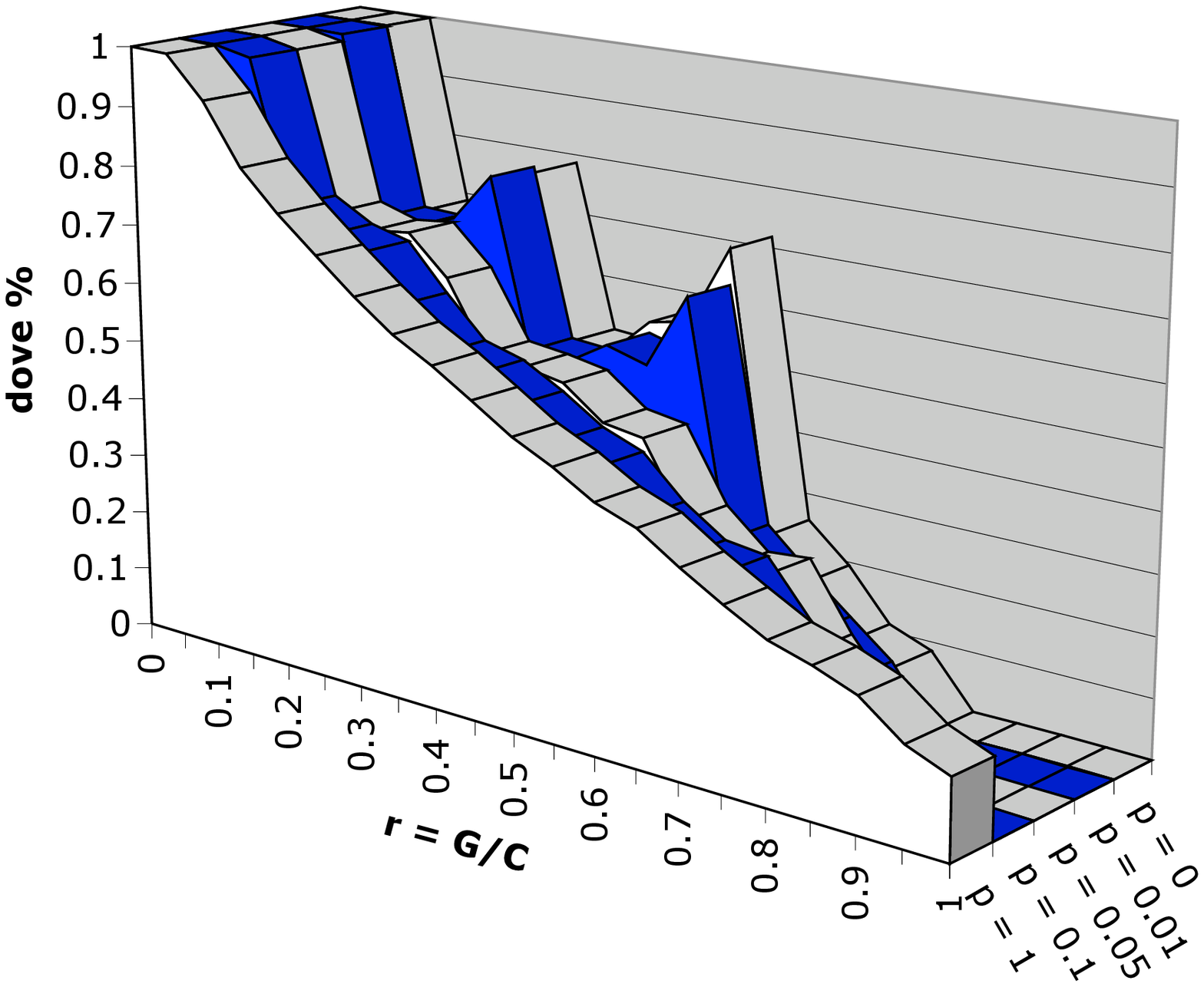}}\protect & \hspace*{1cm} &
\mbox{\includegraphics[width=5.5cm, height=5.5cm]{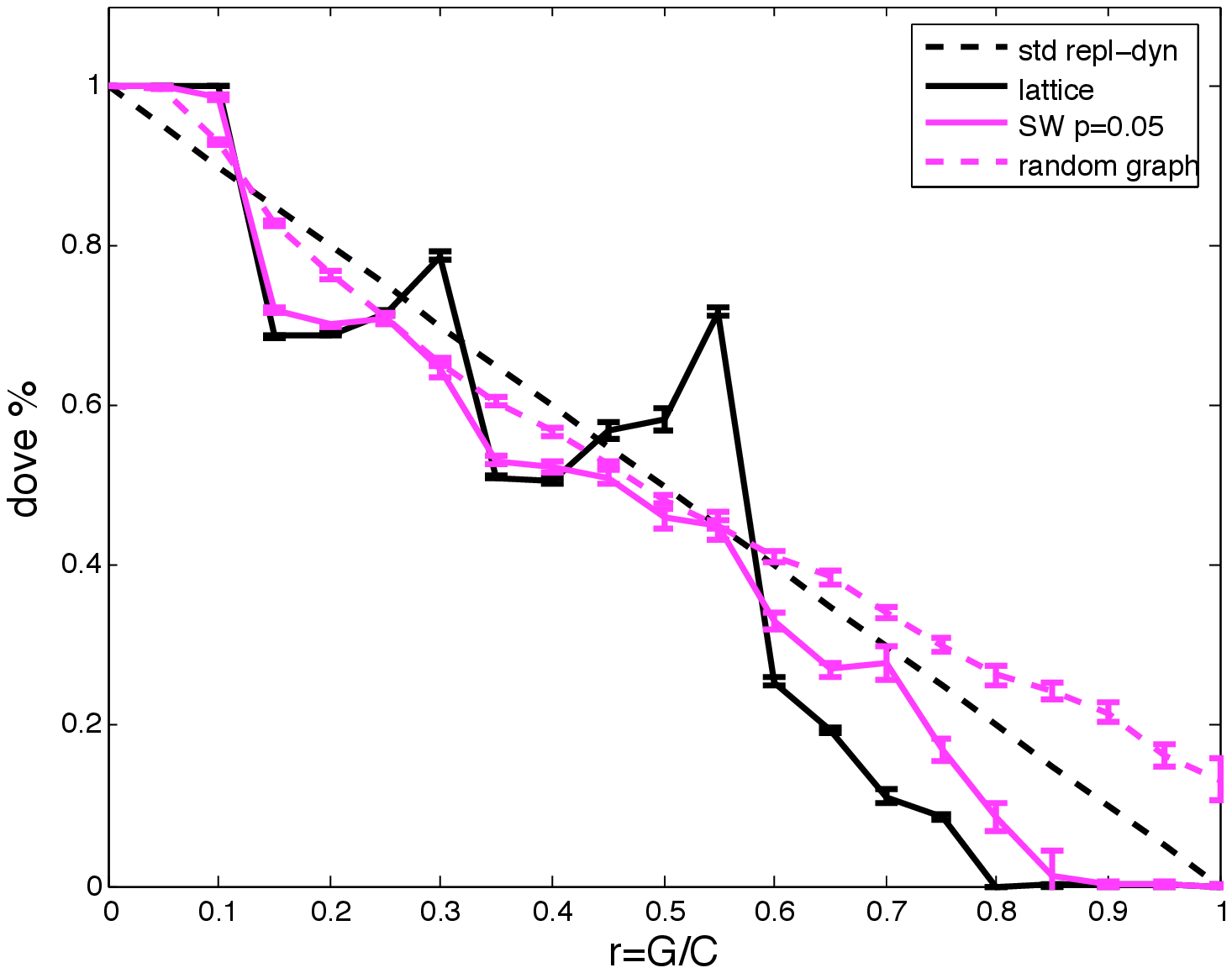}}\\
(a) & & (b)
\end{tabular}
\end{center}
\caption{\label{bto_async}(Color online) asynchronous best-takes-over updating;
(a) frequency of doves as a function of the gain-to-cost ratio $r$ for differents topologies:
lattice ($p=0$), small worlds ($p=0.01$, $p=0.05$, $p=0.1$), random graph ($p=1$);
(b) small world with $p = 0.05$ compared to the grid ($p=0$) and random graph ($p=1$) cases.
Bars indicate standard deviations and the diagonal dashed line is $1-r$ (see text).}
\end{figure}

\begin{figure}[!ht]
\begin{center}
\begin{tabular}{ccc}
\mbox{\includegraphics[width=5.5cm, height=5.5cm]{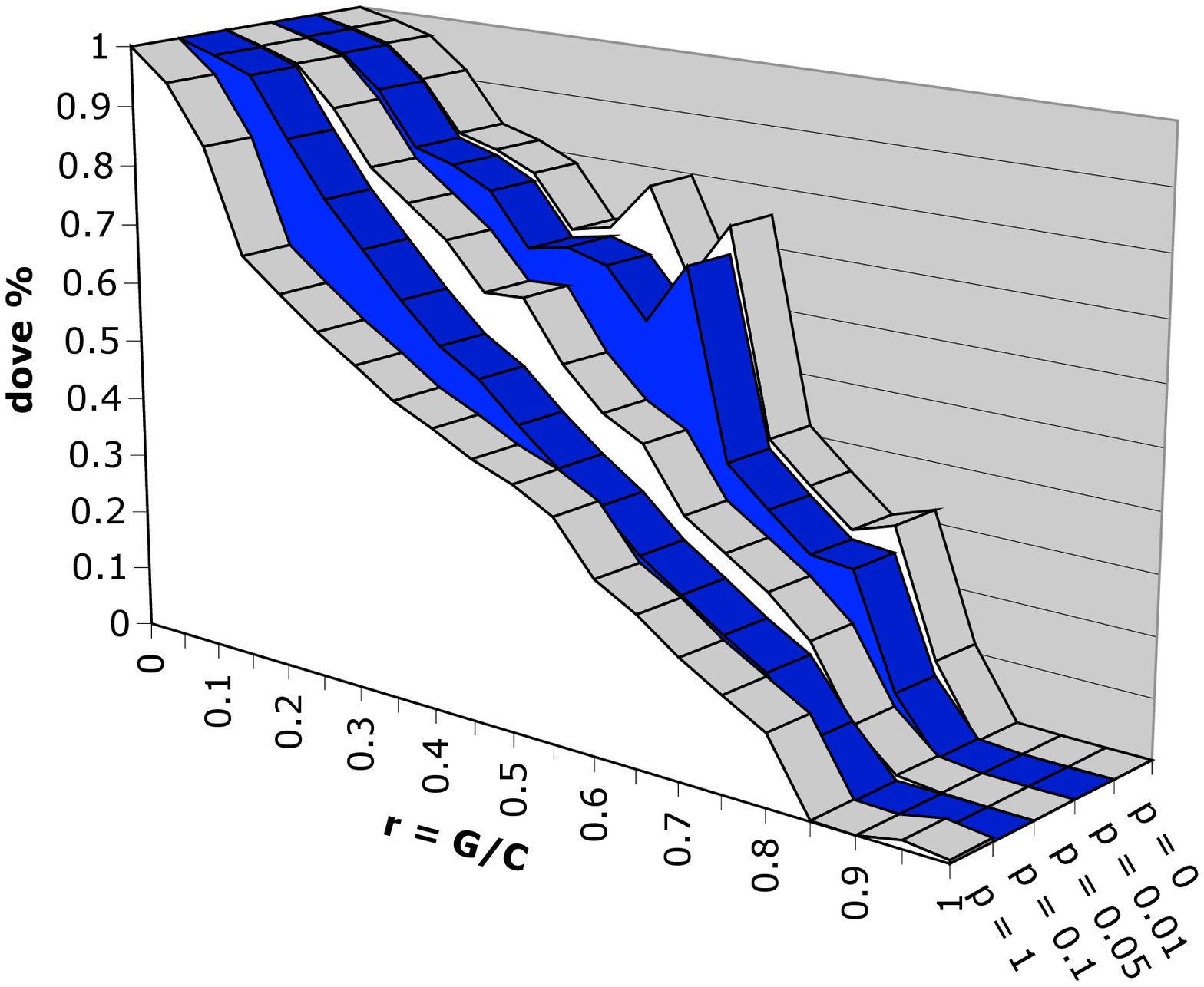}}\protect & \hspace*{1cm} &
\mbox{\includegraphics[width=5.5cm, height=5.5cm]{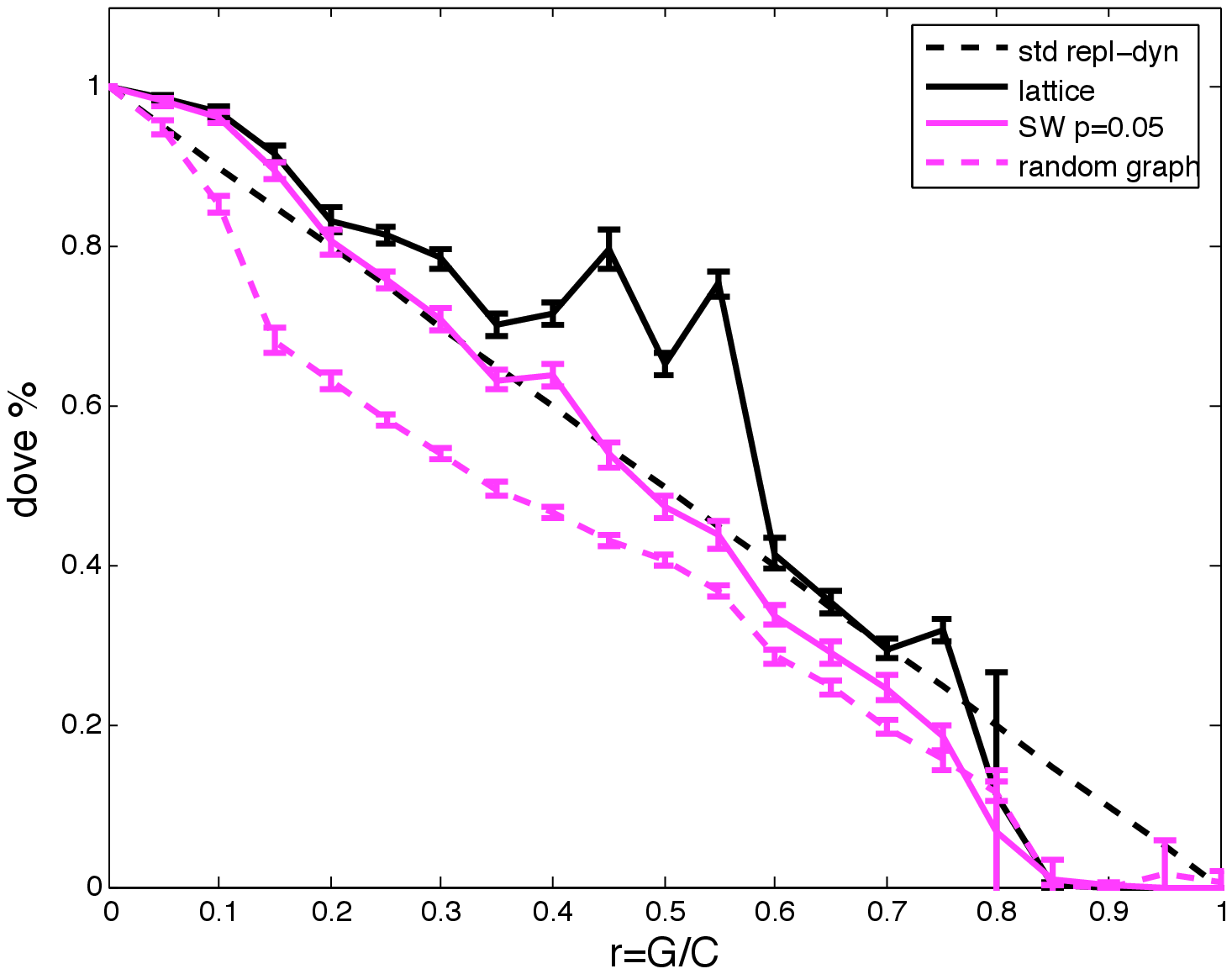}}\\
(a) & & (b)
\end{tabular}
\end{center}
\caption{\label{bto_sync}(Color online) synchronous best-takes-over updating;
(a) frequency of doves as a function of the gain-to-cost ratio $r$ for differents topologies:
lattice ($p=0$), small worlds ($p=0.01$, $p=0.05$, $p=0.1$), random graph ($p=1$);
(b) small world with $p = 0.05$ compared to the grid ($p=0$) and random graph ($p=1$) cases.
Bars indicate standard deviations and the diagonal dashed line is $1-r$ (see text).}
\end{figure}

As pointed out by Hauert and Doebeli \cite{hauer-doeb-2004}, the best-takes-over rule lacks stochasticity,
which in figures \ref{bto_async} and \ref{bto_sync}, translates into discontinuous jumps.

Note that when updating synchronously, best-takes-over is the only rule,
out of the three studied here, where spatial structure actually favors cooperation, as remarked
in \cite{KD-96}, where this was the local update rule used. In fact, the same qualitative results
were found in \cite{hauer-doeb-2004}; however, they appear in the "supplementary material" section,
not in the main text.

\subsection{Time Evolution}
\label{tev}

While the figures in the previous subsections summarize the results at system stability,
here we describe the dynamical behavior of populations through the first $100$ time
steps, where fluctuations might influence the system dynamics.

We have studied both asynchronous and synchronous dynamics for
the three update rules in three topologies each: lattice, random graph, and a small
world with $p=0.05$. This was done for $r=0.7$, where defection predominates. The results are 
relatively uninteresting for the replicator and proportional updates in all topologies. One
observes in the average a monotone decrease of cooperation starting with $50\%$ at time $0$ until
the curve flattens out at the values reported in figures \ref{repl_dyn_async} to \ref{prop_sync}.
The only difference is that
the variance is more pronounced in the proportional case, as one would expect looking
at standard deviations in figures \ref{repl_dyn_async} to \ref{prop_sync}.

\begin{figure}[!ht]
\begin{center}
\begin{tabular}{ccccc}
\mbox{\includegraphics[width=5.5cm, height=5.5cm]{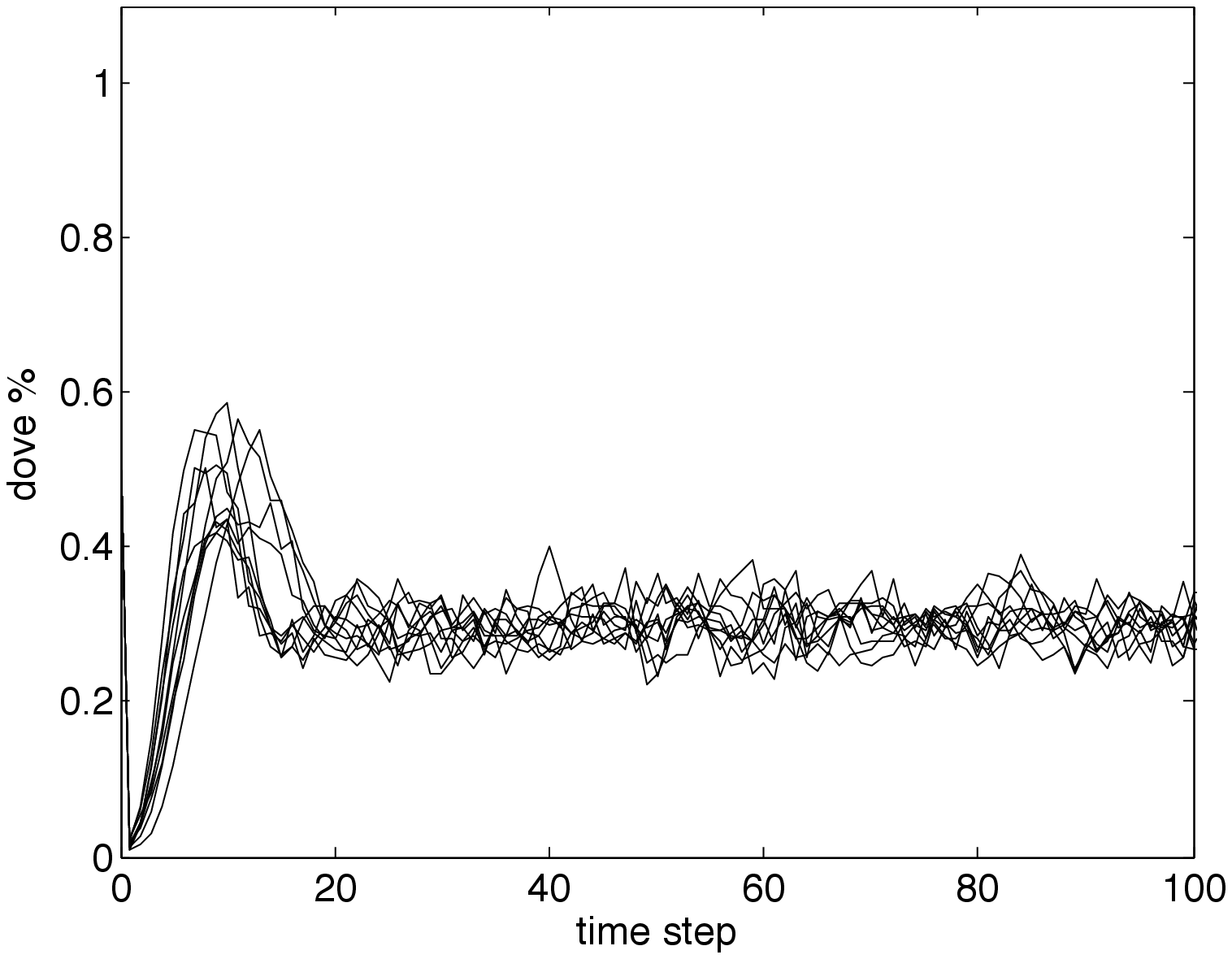}}\protect &
\hspace*{0.3cm} &
\mbox{\includegraphics[width=5.5cm, height=5.5cm]{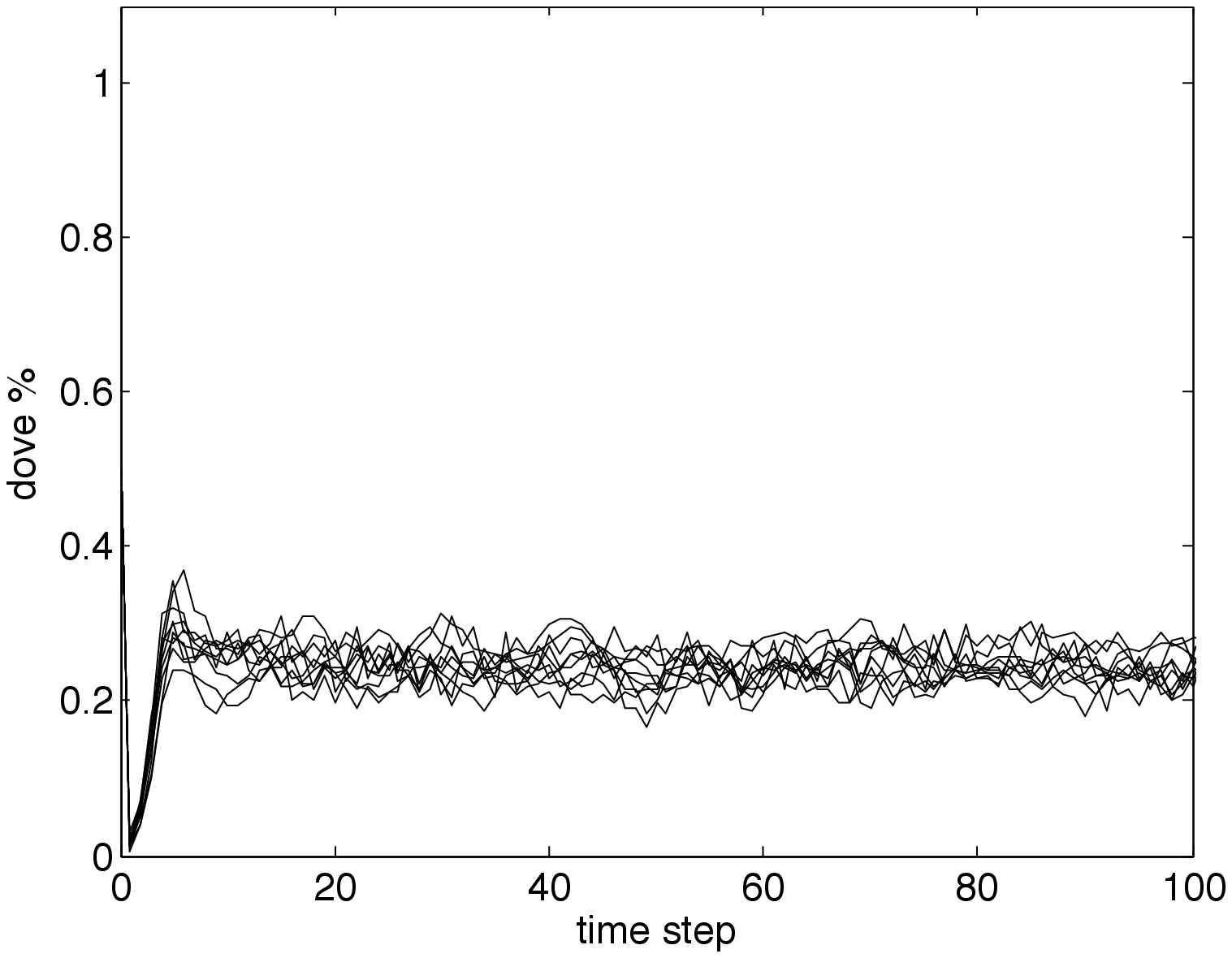}}\protect &
\hspace*{0.3cm} &
\mbox{\includegraphics[width=5.5cm, height=5.5cm]{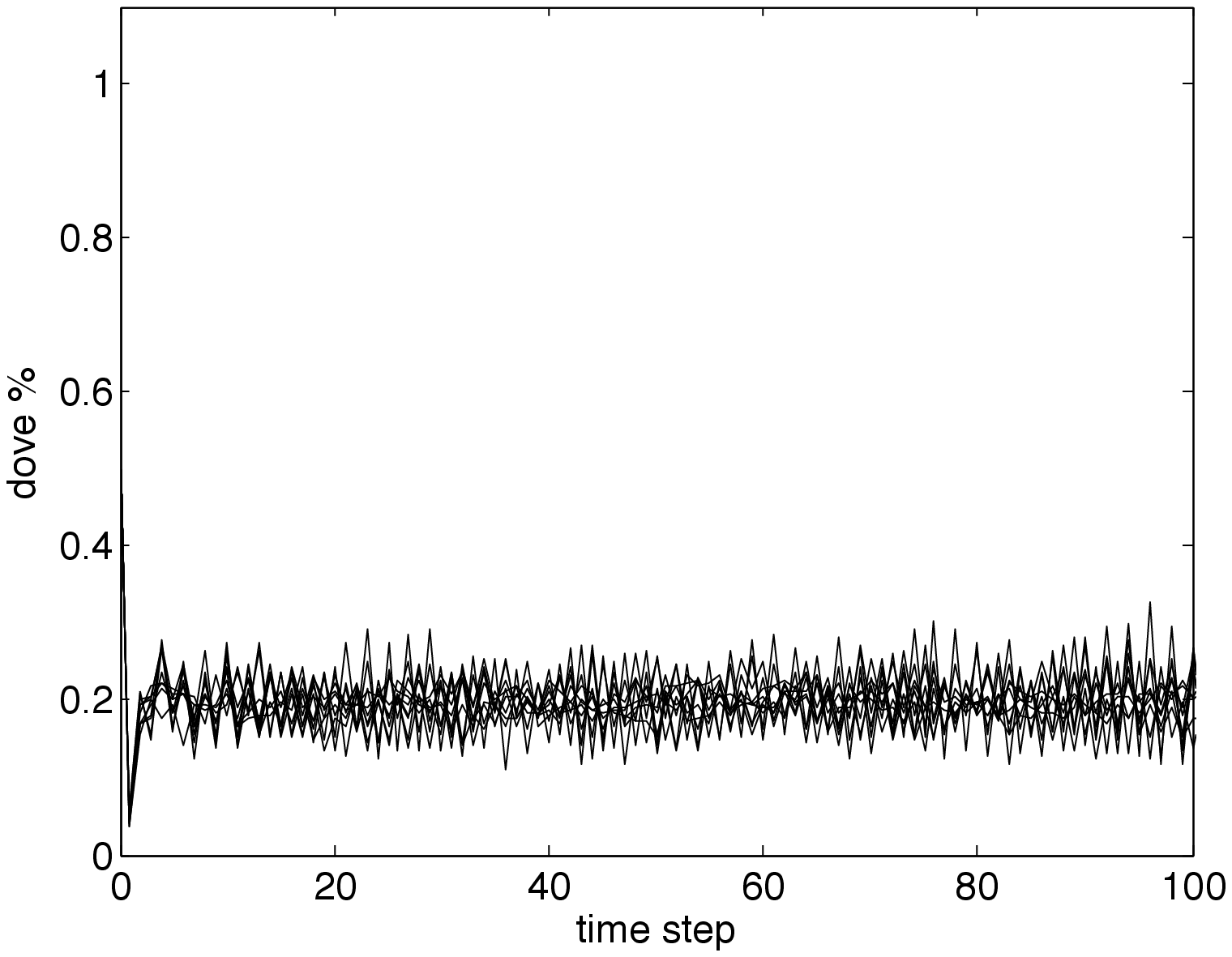}}\\
(a) & & (b) & & (c)
\end{tabular}
\end{center}
\caption{\label{time-ev} time evolution (first $100$ steps) of the proportion of doves
for best-takes-over update; synchronous evolution with $r=0.7$. (a) lattice structure;
(b) small world with $p=0.05$; (c) random graph. Ten randomly chosen evolutions are shown in
each case.}
\end{figure}

The situation is different, and more interesting, in the case of best-takes-over
update whose determinism causes stronger variations. The most striking feature is a 
sudden drop of cooperation at the beginning
of the simulation, followed by an increase and by fluctuations whose amplitude diminishes
over time. The effect is much more pronounced with synchronous dynamics, shown in Fig. \ref{time-ev}, than
with the asynchronous one. The behavior appears in all three topologies but the drop is stronger
in lattices and small worlds with respect to the random graph at earlier times. As time
goes by, fluctuations remain larger in the random graph case. Nevertheless, no experiment
led to total extinction of cooperators at $r=0.7$.

\section{\label{disc}Analysis and Discussion}
If we take a closer look when comparing Fig. \ref{prop_async} and Fig. \ref{prop_sync},
we notice that, for proportional dynamics, asynchronous updating allows for better cooperation than its
synchronous counterpart. The reason for this difference can be intuitively understood
in the following manner:
when updating asynchronously, let us suppose a player $y$ has just imitated the strategy
of one of its neighbors $x$. Another way of viewing this change, is to say that player $x$ has ``infected''
individual $y$ with its strategy. If $x$ is a dove player, making $y$ a dove as well,
not only does the percentage of doves increase in the population, but
the next time either $x$ or $y$ is evaluated for an update, it will be able to take advantage of the other one's
presence to help increase its payoff. Hence, the two players mutually reinforce each other.
Meanwhile, if $y$ is infected by $x$ and turns into a hawk, on the one hand $x$ has successfully propagated
his strategy thus increasing the overall amount of hawks in the population,
but on the other hand this propagation will cause him to have a lower payoff than he previously had.
Not only is $x$'s payoff negatively affected, but $x$'s prescence also harms $y$'s payoff.

The same reasoning cannot be held when updating synchronously.
Indeed, a player $x$ may change strategies
at the same time it infects its neighbor $y$.
So if $x$'s initial strategy was $D$, it might switch
to $H$ as it infects its neighbor $y$, in which case
$x$ will no longer have a positive effect on $y$'s payoff
contrary to what happens in asychronous updating.

When applying the replicators dynamics rule, the small drop of the percentage of doves  seen
on the very left of figures \ref{repl_dyn_async} and \ref{repl_dyn_sync} is due to the fact that for
$r=0$ the game is somewhat degenerated. Indeed, any cluster of more than one hawk will either reduce to
a single hawk or totally disappear, since a dove, no matter what its neighborhood comprises,
will always have a gain of zero while a hawk that interacts with at least one other hawk will have a negative payoff.
The remaining lone hawks will however survive but will not be able to propagate (having a gain 
exactly equal to that of their neighboring doves). The system is thus found locked in a configuration
of a very high proportion of doves with a significant number of isolated hawks.

If $r > 0$, lone hawks always have a higher payoff than the doves in their surroundings and will thus infect one
of its neighbors with its strategy. However for $0 < r \leq 0.1 $, once the pair of hawks is established, their payoff
is lower than the one of any of the doves connected to either one them. Even a dove that interacts with both
hawks has an average payoff still greater than what a hawk composing the
pair receives.
Consequently, when $0 < r \leq 0.1$, clusters of hawks first start by either disappearing
or reducing to single hawks like previously explained for the $r=0$ case, but then these lone hawks
will become pairs of hawks.
If the updates are done synchronously, a pair of hawks will either vanish
or reduce back to a single hawk. One can clearly see that in the long run, hawks will become extinct.
Now if the updates are done asynchronously, a pair cannot totally disappear since only one
player is updated at a time. However, this mechanism of  a pair reducing to a single hawk and
turning back to a pair again
will cause the small groups of two hawks to move across the network and ``collide'' with each other
forming larger groups that reduce back to a single-pair hawk formation. Therefore, after a large
number of time steps, only a very few hawks will survive.  

\begin{figure}[!ht]
\begin{center}
\begin{tabular}{ccc}
\mbox{\includegraphics[width=3.5cm, height=3.5cm]{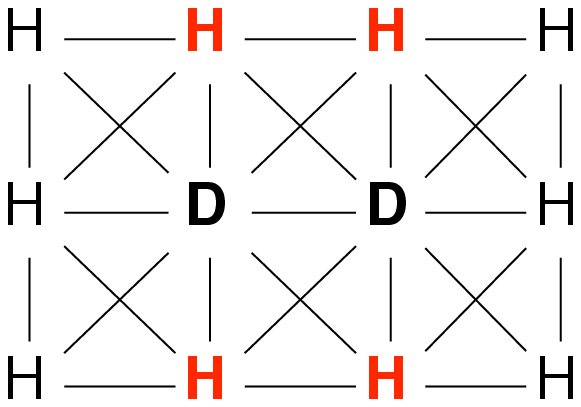}}\protect
&  \hspace{1cm} & \mbox{\includegraphics[width=3.5cm, height=3.5cm]{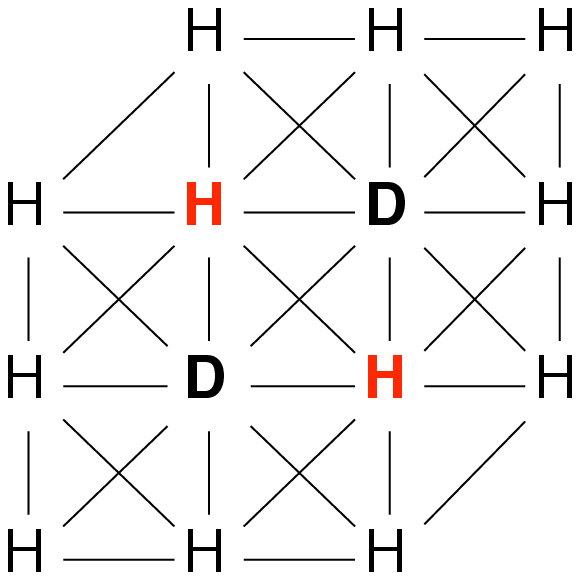}}\\
(a) & \hspace{1cm} & (b)\\
\end{tabular}
\end{center}
\caption{\label{2D_lattice_configs}(Color online) lattice: two possible configurations.}
\end{figure}

If we take another look at figures \ref{repl_dyn_async} and \ref{repl_dyn_sync},
we notice that when the population of players is constrained to a lattice-like structure,
the proportion of doves is reduced to zero for values of
the gain-to-cost ratio greater or equal to approximately $0.8$,
while this not the case when the topology is a random graph.
Let us try to give a qualitative explanation of the two different behaviors:
the first thing to be pointed out is that in the case of the replicators dynamics,
if a dove is surrounded by 8 hawk-neighbors,
it is condemned to die for values of $r$ greater than $\frac{7}{9}$ whatever the topology may be.
However, this does not explain why for these same values, doves no longer exist on
square lattices or small worlds but are able to survive on random graphs.
If the population were mixing, $r=0.8$ would induce a proportion of doves equal to $20\%$.
Therefore, let us suppose that at a certain time step,
there is approximately $20\%$ of doves in our population.
Furthermore, as pointed out by Hauert and Doebeli \cite{hauer-doeb-2004}, in the Hawk-Dove
game on lattices, the doves are usually spread out and form many small-isolated patches.
Thus, we will also suppose $20\%$ of doves in the  population
implies that in a set of players comprising an individual and its immediate eight neighbors,
there are about two doves.
Hence, a D-player has on average one dove and seven hawks in its neighborhood.
In the lattice network, this pair of doves can be linked in two different manners
(see Fig. \ref{2D_lattice_configs}), having either two or four common neighbors,
thus an average of three.

More generally, if we denote $\Gamma$ the clustering coefficient of the graph and $\overline{k}$
the average degree, a pair of doves will have on average $\Gamma(\overline{k} - 1)$
common neighbors.
Let us denote $x$ one of the two doves composing the pair,
$H_x$ a hawk linked to $x$ but not to the other dove of the pair and $H_{x,y}$ one that is connected to both doves.
If $\frac{2}{3} < r < \frac{7}{8}$ and assuming that the hawks surrounding the pair of doves are not interacting with any other doves
(this gives the pair of doves a maximum chance of survival), we have
$$G_{H_x} < G_x < G_{H_{x,y}},$$
where $G_\alpha$ is the average payoff of player $\alpha$

Consequently, according to Eq. (\ref{repl_dyn_eq}),
$x$ can infect $H_x$, and $H_{x,y}$ can infect $x$.

Let us now calculate for what values of $r$ the probability that $x$ invades the site
of at least one $H_x$ is less than an $H_{x,y}$ infecting $x$.
To do so, let us distinguish the case of the asynchronous udating policy from the synchronous one.

\subsubsection*{Asynchronous Dynamics}
The probability that an $H_x$ neighbor is chosen to be updated and adopts strategy $D$ is given by
\begin{equation}
\underbrace{\frac{(1 - \Gamma)(\overline{k} - 1)}{N}}_{(\ast)} \underbrace{\frac{1}{\;\overline{k}\;}}_{(\ast\ast)} \;\phi(G_x - G_{H_x}),
\label{H2D_async}
\end{equation}
where $N$ is the size of the population, $(\ast)$ the probability an $H_x$ hawk is chosen to be updated (among the $N$ players),
$(\ast\ast)$ the probability the chosen $H_x$ hawk compares its payoff with player $x$,
and finally $\phi$ is the function defined in Eq. (\ref{repl_dyn_eq}).

The probability that $x$ is chosen to be updated and is infected by one of the $H_{x,y}$ hawks is given by
\begin{equation}
\underbrace{\frac{1}{N}}_{(\ast)} \underbrace{\frac{\Gamma(\overline{k}-1)}{\overline{k}}}_{(\ast\ast)} \;\phi(G_{H_{x,y}} - G_x),
\label{D2H_async} 
\end{equation}
where $(\ast)$ is the probability $x$ is chosen to be updated,
$(\ast\ast)$ the probability it measures itself against an $H_{x,y}$ neighbor,
and $\phi$ the function defined by Eq. (\ref{repl_dyn_eq}).

For a square lattice with a Moore neighborhood ($\Gamma = \frac{3}{7}$ and $\overline{k} = 8$),
expressions \ref{H2D_async} and \ref{D2H_async} give us $r > \frac{46}{59} \approx 0.78$,
whereas for a random graph, $\Gamma  = \frac{\overline{k}}{N-1} = \frac{8}{2499} \simeq 0.003 \approx 0$ implies that
a pair of doves does not have any common hawk neighbors enabling them to survive
if $r < \frac{7}{8}$.
As for the small-world cases, the clustering coefficient is very close to that of the lattice, generating a behavior
pratically identical to the latter.
This gives a qualitative explanation for the difference observed in Fig. \ref{repl_dyn_async}.

\subsubsection*{Synchronous Dynamics}
The probability that at least one $H_x$ adopts strategy $D$ is given by
\begin{equation}
1 - \overbrace{[1 - \underbrace{\frac{1}{\;\overline{k}\;}\;\phi(G_x - G_{H_x})}_{(\ast)}]^{(1-\Gamma)(\overline{k}-1)}}^{(\ast\ast)},
\label{H2D_sync}
\end{equation}
where $(\ast)$ is the probability a specific $H_{x}$ turns into a dove and $(\ast\ast)$ the probability none of the $H_x$
adopt strategy $D$.

The probability that $x$ adopts the hawk strategy is given by
\begin{equation}
\underbrace{\frac{\Gamma(\overline{k} - 1)}{\overline{k}}}_{(\ast)} \;\phi(G_{H_{x,y}} - G_x),
\label{D2H_sync}
\end{equation}
where $(\ast)$ is the probability player $x$ compares its payoff with one of its $H_{x,y}$ neighbors.

For a square lattice with a Moore neighborhood ($\Gamma = \frac{3}{7}$ and $\overline{k} = 8$),
expressions \ref{H2D_sync} and \ref{D2H_sync} yield
$$
1 - \left[1 - \frac{1}{8}\left(\frac{-8G+7C}{G+C}\right)\right]^4 < \frac{3}{8}\left(\frac{9G - 6C}{G+C}\right),
$$
and given that $\frac{G}{C} = r$, we obtain
$$
1 - \left[1 - \frac{1}{8}\left(\frac{-8r+7}{r+1}\right)\right]^4 < \frac{3}{8}\left(\frac{9r - 6}{r+1}\right),
$$
which is true for about $r > 0.775$. This also holds for the small-world cases, since, once again,
they have a $\Gamma$ close to the one of the lattice. 

For a random graph of $N=2500$ nodes and $\overline{k}=8$, we have $\Gamma \approx 0$.
Therefore, a pair of doves has a negligible probability of having a hawk neighbor in common and thus cannot be infected by the H strategy if $r < \frac{7}{8}$.
This enables a small percentage of doves to survive on the random graph topology contrary to the lattice and small-world
networks (see Fig. \ref{repl_dyn_sync}). 

In a few words, whether the update policy is asynchronous or synchronous, as soon as $r > \frac{7}{9}$,
isolated doves, as well as pairs of doves surrounded by hawks, will end up disappearing in the
lattice and small-world cases due to the high clustering coefficient.
However, in the random graph scenario, although isolated doves are also bound to die if $r > \frac{7}{9}$,
pairs of doves have a more than even chance of surviving (at least as long as $r < \frac{7}{8}$).

\section{\label{concl}Conclusions}
In this work we clarify previous partially contradictory results on cooperation in populations playing the
Hawk-Dove game on regular grids. Furthermore, we notably extend the study to Watts--Strogatz small-world graphs, as
these population structures lie between the two extreme cases of regular lattices and random graphs,
and are a first simple step towards real social interaction networks. This allows us to
unravel the role of network clustering on cooperation in the Hawk-Dove game.
We find that, in general, spatial structure on the network of interactions in the  game
either favors or inhibits cooperation with respect to the perfectly mixed case.
The influence it has depends not only on the rule that determines a player's future strategy,
but also on the value of the gain-to-cost ratio $G/C$ and to a lesser degree,
on the synchronous and asynchronous timing of events.

In the case of the best-takes-over rule, dove-like behavior is advantaged if synchronous
update is used but the rule is noisy due to its discrete nature.
In the case of the proportional update
rule, giving the network a regular structure tends to increase
the percentage of the strategy that would already be in majority on a random graph
configuration of the population. The more important the structure, in terms of clustering
coefficient, the higher the percentage of the dominant strategy. In fact, cooperation predominates
for low to medium $r$ values, while for higher $r$ values cooperation falls below the 
large population, mixing case.
Finally, the replicators dynamics rule tends to favor hawks over doves on
spatially structured topologies such as small worlds and square lattices, thus
confirming previous results for regular lattices and extending them to small-world networks.
In the end, although small-world topologies show behaviors that are somewhat in between
those of the random graph and the two-dimensional lattice, they usually tend more
towards the latter, at least in terms of cooperation level.

In this work, we have used static
network structures, which is a useful step but it is not realistic enough as the interactions
themselves help shape the network. In future work we shall
extend the study using more faithful social network structures, including their dynamical aspects.

\begin{small}
\bibliographystyle{plain}
\bibliography{games}

\begin{thebibliography}{23}
\expandafter\ifx\csname natexlab\endcsname\relax\def\natexlab#1{#1}\fi
\expandafter\ifx\csname bibnamefont\endcsname\relax
  \def\bibnamefont#1{#1}\fi
\expandafter\ifx\csname bibfnamefont\endcsname\relax
  \def\bibfnamefont#1{#1}\fi
\expandafter\ifx\csname citenamefont\endcsname\relax
  \def\citenamefont#1{#1}\fi
\expandafter\ifx\csname url\endcsname\relax
  \def\url#1{\texttt{#1}}\fi
\expandafter\ifx\csname urlprefix\endcsname\relax\def\urlprefix{URL }\fi
\providecommand{\bibinfo}[2]{#2}
\providecommand{\eprint}[2][]{\url{#2}}

\bibitem[{\citenamefont{Axelrod}(1984)}]{axe84}
\bibinfo{author}{\bibfnamefont{R.}~\bibnamefont{Axelrod}},
  \emph{\bibinfo{title}{The Evolution of Cooperation}}
  (\bibinfo{publisher}{Basic Books, Inc.}, \bibinfo{address}{New-York},
  \bibinfo{year}{1984}).

\bibitem[{\citenamefont{Poundstone}(1992)}]{poundstone92}
\bibinfo{author}{\bibfnamefont{W.}~\bibnamefont{Poundstone}},
  \emph{\bibinfo{title}{The Prisoner's Dilemma}}
  (\bibinfo{publisher}{Doubleday}, \bibinfo{address}{New York},
  \bibinfo{year}{1992}).

\bibitem[{\citenamefont{Hofbauer and Sigmund}(1998)}]{hofb-sigm-book-98}
\bibinfo{author}{\bibfnamefont{J.}~\bibnamefont{Hofbauer}} \bibnamefont{and}
  \bibinfo{author}{\bibfnamefont{K.}~\bibnamefont{Sigmund}},
  \emph{\bibinfo{title}{Evolutionary Games and Population Dynamics}}
  (\bibinfo{publisher}{Cambridge University Press, Cambridge, UK},
  \bibinfo{year}{1998}).

\bibitem[{\citenamefont{Nowak and May}(1992)}]{nowakmay92}
\bibinfo{author}{\bibfnamefont{M.~A.} \bibnamefont{Nowak}} \bibnamefont{and}
  \bibinfo{author}{\bibfnamefont{R.~M.} \bibnamefont{May}},
  \bibinfo{journal}{Nature} \textbf{\bibinfo{volume}{359}},
  \bibinfo{pages}{826} (\bibinfo{year}{1992}).

\bibitem[{\citenamefont{Killingback and Doebeli}(1996)}]{KD-96}
\bibinfo{author}{\bibfnamefont{T.}~\bibnamefont{Killingback}} \bibnamefont{and}
  \bibinfo{author}{\bibfnamefont{M.}~\bibnamefont{Doebeli}},
  \bibinfo{journal}{Proceedings of the Royal Society of London B}
  \textbf{\bibinfo{volume}{263}}, \bibinfo{pages}{1135} (\bibinfo{year}{1996}).

\bibitem[{\citenamefont{Hauert and Doebeli}(2004)}]{hauer-doeb-2004}
\bibinfo{author}{\bibfnamefont{C.}~\bibnamefont{Hauert}} \bibnamefont{and}
  \bibinfo{author}{\bibfnamefont{M.}~\bibnamefont{Doebeli}},
  \bibinfo{journal}{Nature} \textbf{\bibinfo{volume}{428}},
  \bibinfo{pages}{643} (\bibinfo{year}{2004}).

\bibitem[{\citenamefont{Sysi-Aho et~al.}(2005)\citenamefont{Sysi-Aho,
  Saram\"aki, Kert\'esz, and Kaski}}]{myopic-hd-05}
\bibinfo{author}{\bibfnamefont{M.}~\bibnamefont{Sysi-Aho}},
  \bibinfo{author}{\bibfnamefont{J.}~\bibnamefont{Saram\"aki}},
  \bibinfo{author}{\bibfnamefont{J.}~\bibnamefont{Kert\'esz}},
  \bibnamefont{and} \bibinfo{author}{\bibfnamefont{K.}~\bibnamefont{Kaski}},
  \bibinfo{journal}{Eur. Phys. Jour. B} \textbf{\bibinfo{volume}{44}},
  \bibinfo{pages}{129} (\bibinfo{year}{2005}).

\bibitem[{\citenamefont{Milgram}(1967)}]{milgram67}
\bibinfo{author}{\bibfnamefont{S.}~\bibnamefont{Milgram}},
  \bibinfo{journal}{Psychology Today} \textbf{\bibinfo{volume}{2}},
  \bibinfo{pages}{60} (\bibinfo{year}{1967}).

\bibitem[{\citenamefont{Newman}(2003)}]{newman-03}
\bibinfo{author}{\bibfnamefont{M.~E.~J.} \bibnamefont{Newman}},
  \bibinfo{journal}{{SIAM} Review} \textbf{\bibinfo{volume}{45}},
  \bibinfo{pages}{167} (\bibinfo{year}{2003}).

\bibitem[{\citenamefont{Watts and Strogatz}(1998)}]{watts-strogatz-98}
\bibinfo{author}{\bibfnamefont{D.~J.} \bibnamefont{Watts}} \bibnamefont{and}
  \bibinfo{author}{\bibfnamefont{S.~H.} \bibnamefont{Strogatz}},
  \bibinfo{journal}{Nature} \textbf{\bibinfo{volume}{393}},
  \bibinfo{pages}{440} (\bibinfo{year}{1998}).

\bibitem[{\citenamefont{Abramson and Kuperman}(2001)}]{social-pd-kup-01}
\bibinfo{author}{\bibfnamefont{G.}~\bibnamefont{Abramson}} \bibnamefont{and}
  \bibinfo{author}{\bibfnamefont{M.}~\bibnamefont{Kuperman}},
  \bibinfo{journal}{Phys. Rev. E} \textbf{\bibinfo{volume}{63}},
  \bibinfo{pages}{030901} (\bibinfo{year}{2001}).

\bibitem[{\citenamefont{Kim et~al.}(2002)\citenamefont{Kim, Trusina, Holme,
  Minnhagen, Chung, and Choi}}]{pd-dyn-sw-02}
\bibinfo{author}{\bibfnamefont{B.~J.} \bibnamefont{Kim}},
  \bibinfo{author}{\bibfnamefont{A.}~\bibnamefont{Trusina}},
  \bibinfo{author}{\bibfnamefont{P.}~\bibnamefont{Holme}},
  \bibinfo{author}{\bibfnamefont{P.}~\bibnamefont{Minnhagen}},
  \bibinfo{author}{\bibfnamefont{J.~S.} \bibnamefont{Chung}}, \bibnamefont{and}
  \bibinfo{author}{\bibfnamefont{M.~Y.} \bibnamefont{Choi}},
  \bibinfo{journal}{Phys. Rev. E} \textbf{\bibinfo{volume}{66}},
  \bibinfo{pages}{021907} (\bibinfo{year}{2002}).

\bibitem[{\citenamefont{Watts}(1999)}]{watts99}
\bibinfo{author}{\bibfnamefont{D.~J.} \bibnamefont{Watts}},
  \emph{\bibinfo{title}{Small worlds: The Dynamics of Networks between Order
  and Randomness}} (\bibinfo{publisher}{Princeton University Press},
  \bibinfo{address}{Princeton NJ}, \bibinfo{year}{1999}).

\bibitem[{\citenamefont{Santos and Pacheco}(2005)}]{santos-pach-05}
\bibinfo{author}{\bibfnamefont{F.~C.} \bibnamefont{Santos}} \bibnamefont{and}
  \bibinfo{author}{\bibfnamefont{J.~M.} \bibnamefont{Pacheco}},
  \bibinfo{journal}{Phys. Rev. Lett.} \textbf{\bibinfo{volume}{95}},
  \bibinfo{pages}{098104} (\bibinfo{year}{2005}).

\bibitem[{\citenamefont{Ebel et~al.}(2003)\citenamefont{Ebel, Davidsen, and
  Bornholdt}}]{ebel-dav-born-03}
\bibinfo{author}{\bibfnamefont{H.}~\bibnamefont{Ebel}},
  \bibinfo{author}{\bibfnamefont{J.}~\bibnamefont{Davidsen}}, \bibnamefont{and}
  \bibinfo{author}{\bibfnamefont{S.}~\bibnamefont{Bornholdt}},
  \bibinfo{journal}{Complexity} \textbf{\bibinfo{volume}{8}},
  \bibinfo{pages}{24} (\bibinfo{year}{2003}).

\bibitem[{\citenamefont{Jin et~al.}(2001)\citenamefont{Jin, Girvan, and
  Newman}}]{jin-gir-newman-01}
\bibinfo{author}{\bibfnamefont{E.~M.} \bibnamefont{Jin}},
  \bibinfo{author}{\bibfnamefont{M.}~\bibnamefont{Girvan}}, \bibnamefont{and}
  \bibinfo{author}{\bibfnamefont{M.~E.~J.} \bibnamefont{Newman}},
  \bibinfo{journal}{Phys. Rev. E} \textbf{\bibinfo{volume}{64}},
  \bibinfo{pages}{046132} (\bibinfo{year}{2001}).

\bibitem[{\citenamefont{Boccara}(2004)}]{boccara-04}
\bibinfo{author}{\bibfnamefont{N.}~\bibnamefont{Boccara}},
  \emph{\bibinfo{title}{Modeling Complex Systems}}
  (\bibinfo{publisher}{Springer, New York, Berlin, Heidelberg},
  \bibinfo{year}{2004}).

\bibitem[{\citenamefont{Luthi et~al.}(2005{\natexlab{a}})\citenamefont{Luthi,
  Giacobini, and Tomassini}}]{games-ecal-05}
\bibinfo{author}{\bibfnamefont{L.}~\bibnamefont{Luthi}},
  \bibinfo{author}{\bibfnamefont{M.}~\bibnamefont{Giacobini}},
  \bibnamefont{and}
  \bibinfo{author}{\bibfnamefont{M.}~\bibnamefont{Tomassini}}, in
  \emph{\bibinfo{booktitle}{Advances in Artificial Life, Eighth European
  Conference, ECAL 2005}}, edited by \bibinfo{editor}{\bibfnamefont{M.~C.}
  \bibnamefont{et~al.}} (\bibinfo{publisher}{Springer, Berlin, Heidelberg, New
  York}, \bibinfo{year}{2005}{\natexlab{a}}), vol. \bibinfo{volume}{3630} of
  \emph{\bibinfo{series}{Lecture Notes in Artificial Intelligence}}, pp.
  \bibinfo{pages}{665--674}.

\bibitem[{\citenamefont{Luthi et~al.}(2005{\natexlab{b}})\citenamefont{Luthi,
  Giacobini, and Tomassini}}]{luthi-giac-tom-05}
\bibinfo{author}{\bibfnamefont{L.}~\bibnamefont{Luthi}},
  \bibinfo{author}{\bibfnamefont{M.}~\bibnamefont{Giacobini}},
  \bibnamefont{and}
  \bibinfo{author}{\bibfnamefont{M.}~\bibnamefont{Tomassini}}, in
  \emph{\bibinfo{booktitle}{Proceedings of the IEEE Symposium on Computational
  Intelligence and Games}}, edited by
  \bibinfo{editor}{\bibfnamefont{G.}~\bibnamefont{Kendall}} \bibnamefont{and}
  \bibinfo{editor}{\bibfnamefont{S.}~\bibnamefont{Lucas}}
  (\bibinfo{publisher}{IEEE Press, Piscataway, NJ},
  \bibinfo{year}{2005}{\natexlab{b}}), pp. \bibinfo{pages}{225--232}.

\bibitem[{\citenamefont{Zimmermann et~al.}(2004)\citenamefont{Zimmermann,
  Egu\'iluz, and Miguel}}]{zimm-et-al-04}
\bibinfo{author}{\bibfnamefont{M.~G.} \bibnamefont{Zimmermann}},
  \bibinfo{author}{\bibfnamefont{V.~M.} \bibnamefont{Egu\'iluz}},
  \bibnamefont{and} \bibinfo{author}{\bibfnamefont{M.~S.}
  \bibnamefont{Miguel}}, \bibinfo{journal}{Phys. Rev. E}
  \textbf{\bibinfo{volume}{69}}, \bibinfo{pages}{065102(R)}
  (\bibinfo{year}{2004}).

\bibitem[{\citenamefont{Nowak et~al.}(1994)\citenamefont{Nowak, Bonhoeffer, and
  May}}]{nowaketal94}
\bibinfo{author}{\bibfnamefont{M.~A.} \bibnamefont{Nowak}},
  \bibinfo{author}{\bibfnamefont{S.}~\bibnamefont{Bonhoeffer}},
  \bibnamefont{and} \bibinfo{author}{\bibfnamefont{R.~M.} \bibnamefont{May}},
  \bibinfo{journal}{Proceedings of the National Academy of Sciences USA}
  \textbf{\bibinfo{volume}{91}}, \bibinfo{pages}{4877} (\bibinfo{year}{1994}).

\bibitem[{\citenamefont{Huberman and Glance}(1993)}]{hubglance93}
\bibinfo{author}{\bibfnamefont{B.~A.} \bibnamefont{Huberman}} \bibnamefont{and}
  \bibinfo{author}{\bibfnamefont{N.~S.} \bibnamefont{Glance}},
  \bibinfo{journal}{Proceedings of the National Academy of Sciences USA}
  \textbf{\bibinfo{volume}{90}}, \bibinfo{pages}{7716} (\bibinfo{year}{1993}).

\bibitem[{\citenamefont{Baalen}(2000)}]{van-baalen-00}
\bibinfo{author}{\bibfnamefont{M.~V.} \bibnamefont{Baalen}}, in
  \emph{\bibinfo{booktitle}{The Geometry of Ecological Interactions:
  Simplifying Spatial Complexity}}, edited by
  \bibinfo{editor}{\bibfnamefont{U.}~\bibnamefont{Dieckmann}},
  \bibinfo{editor}{\bibfnamefont{R.}~\bibnamefont{Law}}, \bibnamefont{and}
  \bibinfo{editor}{\bibfnamefont{J.~A.~J.} \bibnamefont{Metz}}
  (\bibinfo{publisher}{Cambridge University Press, Cambridge, UK},
  \bibinfo{year}{2000}), pp. \bibinfo{pages}{359--387}.

\end{thebibliography}
\end{small}

\end{document}